\documentclass[global,twocolumn]{svjour}
\usepackage{graphics}
\usepackage{amssymb}
\journalname{Applied Physics B}
\begin{document}
\title{Precision spectroscopy with two correlated atoms}
\author{M.~Chwalla\inst{1}
\and K.~Kim\inst{1} \and T.~Monz\inst{1} \and P.~Schindler\inst{1}
\and M.~Riebe \inst{1} \and C.~F.~Roos \inst{1,2}
\and R.~Blatt
\inst{1,2}
}                     
%
%
\institute{Institut f\"ur Experimentalphysik, Universit\"{a}t
Innsbruck, Technikerstra{\ss}e 25, A--6020 Innsbruck, Austria \and
Institut f\"ur Quantenoptik und Quanteninformation der
\"Osterreichischen Akademie der Wissenschaften,
Technikerstra{\ss}e 21a, A--6020 Innsbruck,
Austria\\\email{christian.roos@uibk.ac.at} }
\date{Received: date / Revised version: date}
%
\maketitle
\begin{abstract}
We discuss techniques that allow for long coherence times in laser
spectroscopy experiments with two trapped ions. We show that for
this purpose not only entangled ions prepared in decoherence-free
subspaces can be used but also a pair of ions that are not
entangled but subject to the same kind of phase noise. We apply
this technique to a measurement of the electric quadrupole moment
of the $3d^2D_{5/2}$ state of $^{40}$Ca$^+$ and to a measurement
of the line width of an ultra-stable laser exciting a pair of
$^{40}$Ca$^+$ ions.
\end{abstract}

%

\section{Introduction}
\label{intro}
Single trapped and laser-cooled ions are a nearly ideal system for
high-resolution laser spectroscopy. The ion's internal as well as
external quantum degrees of freedom can be controlled by coherent
laser-atom interactions with high accuracy. At the same time, the
ion is well isolated against detrimental influences of a
decohering environment. The combination of these two properties
have enabled spectacular ion trap experiments aiming at building
better atomic clocks
\cite{Margolis04,Schneider05,Oskay06,Rosenband07}, creating
entangled states \cite{Leibfried05,Haeffner05}, and processing
quantum information \cite{Riebe04,Barrett04}. The creation of
multi-particle entangled atomic states is of interest in the
context of quantum information processing but also for laser
spectroscopy experiments. Maximally entangled states offer the
prospect of improving the signal-to-noise ratio in precision
spectroscopy experiments \cite{Bollinger96,Leibfried04} and they
can be used to read out the quantum state of an atom by entangling
it with another atom that is easier to detect \cite{Schmidt05}.
Furthermore, in the state space of composite quantum systems made
out of two or more atoms, it is sometimes possible to identify
subspaces that are decoherence free \cite{Lidar98} with respect to
the dominant source of noise in the single-atom system
\cite{Kielpinski01}. It has been shown that such decoherence-free
subspaces allow the preparation of entangled states of two ions
with an entanglement lifetime of about one second predominantly
limited by spontaneous decay of the involved metastable states
\cite{Roos04}. These long entanglement lifetimes substantiate the
usefulness of entangled states for precision measurements
\cite{Roos05}. In ref. \cite{Roos06} we applied this measurement
technique to a pair of entangled $^{40}$Ca$^+$ ions for a
precision measurement of the electric quadrupole moment of the
$D_{5/2}$ state. The question arises of whether entanglement is an
indispensable ingredient in these measurements. In this paper, we
show that for a pair of correlated atoms coherence times exceeding
the single-atom coherence times can be obtained even if the atoms
are not entangled. The technique is applicable to measurements of
energy-level shifts and transition frequencies in the presence of
correlated noise. After a discussion of the measurement principle,
we present two experiments applying the technique to a measurement
of an electric quadrupole moment and to a determination of the
line width of a narrow-band laser.

\section{Spectroscopy in decoherence-free subspaces}
In atomic high-resolution spectroscopy, dephasing is often the
most important factor limiting the attainable spectral resolution.
Possible sources of dephasing are fluctuating electromagnetic
fields giving rise to random energy-level shifts but also the
finite spectral line width of probe lasers. Under these
conditions, two atoms located in close proximity to each other are
likely to experience the same kind of noise, i.e. they are subject
to collective decoherence. The collective character of the
decoherence has the important consequence that it does not affect
the entangled two-atom state
\begin{equation}
\Psi_+=\frac{1}{\sqrt{2}}\left(|g\rangle|e\rangle+|e\rangle|g\rangle\right),
\label{Bellstate}
\end{equation}
as both parts of the superposition are shifted by the same amount
of energy by fluctuating fields. Here, for the sake of simplicity,
$|g\rangle$ and $|e\rangle$ denote the ground and excited states
of a two-level atom. Because of its immunity against collective
decoherence, the entangled state $\Psi_+$ is much more robust than
a single-atom superposition state
$\frac{1}{\sqrt{2}}(|g\rangle+|e\rangle)$. This property makes
states like $\Psi_+$ interesting candidates for high-precision
spectroscopy. In the following, we will first discuss how to use
Bell states for the measurement of energy-level shifts. Then, it
will be shown that certain unentangled two-atom states can have
similar advantages over single-atom superposition states albeit at
lower signal-to-noise ratio.

\subsection{Spectroscopy with entangled states}
\label{sec2:entangled} In a Ramsey experiment, spectroscopic
information is inferred from a measurement of the relative phase
$\phi$ of the superposition state
$\frac{1}{\sqrt{2}}(|g\rangle+e^{i\phi}|e\rangle)$. The phase is
measured by mapping the states
$\frac{1}{\sqrt{2}}(|g\rangle\pm|e\rangle)$ to the measurement
basis $\{|g\rangle,|e\rangle\}$ by means of a $\pi/2$ pulse. In
close analogy, spectroscopy with entangled states is based on a
measurement of the relative phase $\phi$ of the Bell state
$\Psi_\phi=\frac{1}{\sqrt{2}}(|g\rangle|e\rangle
+e^{i\phi}|e\rangle|g\rangle)$. Here, the phase is determined by
applying $\pi/2$ pulses to both atoms followed by state detection.
$\pi/2$ pulses with the same laser phase on both atoms map the
singlet state
$\frac{1}{\sqrt{2}}\left(|g\rangle|e\rangle-|e\rangle|g\rangle\right)$
to itself, whereas the triplet state
$\frac{1}{\sqrt{2}}\left(|g\rangle|e\rangle+|e\rangle|g\rangle\right)$
is mapped to a state
$\frac{1}{\sqrt{2}}\left(|g\rangle|g\rangle+e^{i\alpha}|e\rangle|e\rangle\right)$
with different parity. Therefore, measurement of the parity
operator $\sigma_z^{(1)}\sigma_z^{(2)}$ yields information about
the relative phase since
$\langle\sigma_z^{(1)}\sigma_z^{(2)}\rangle=\cos\phi$. If the
atomic transition frequencies are not exactly equal but differ by
an amount $\delta$, the phase will evolve as a function of time
$\tau$ according to $\phi(\tau)=\phi_0 + \delta\tau$. Then,
measurement of the phase-evolution rate provides information about
the difference frequency $\delta$. To keep the notation simple, it
was assumed that in both atoms the same energy levels participated
in the superposition state of eq. (\ref{Bellstate}). In general,
this does not need to be the case and the phase evolution is given
by
$\phi(\tau)=((\omega_{A_1}-\omega_{L_1})\pm(\omega_{A_2}-\omega_{L_2}))\tau$.
Here, $\omega_{A_{1,2}}$ denote the atomic transition frequencies
of atom 1 and atom 2, and $\omega_{L_{1,2}}$ are the laser
frequencies used for exciting the corresponding transitions. The
minus sign applies if in the Bell state the ground state of atom 1
is associated with an excited state of atom 2 and vice versa. If
the Bell state is a superposition of both atoms being in the
ground state or both in the excited state, the plus sign is
appropriate.

\subsection{Spectroscopy with unentangled states of two atoms}
\label{sec2:unentangled} One may wonder whether entanglement is
absolutely necessary for observing long coherence times in
experiments with two atoms. In fact, it turns out that the kind of
measurement outlined above is applicable even to completely
unentangled atoms. If the atoms are initially prepared in the
product state
\begin{eqnarray}
\Psi_p &=& \frac{1}{2}(|g\rangle+|e\rangle)\otimes(|g\rangle+|e\rangle)\\
&=&\frac{1}{\sqrt{2}}\Psi_++\frac{1}{2}|g\rangle|g\rangle+\frac{1}{2}|e\rangle|e\rangle,\nonumber
\label{productstate}
\end{eqnarray}
this state will quickly dephase under the influence of collective
phase noise. The resulting mixed state
\begin{equation}
\rho_p =
\frac{1}{2}|\Psi_+\rangle\langle\Psi_+|+\frac{1}{4}|gg\rangle\langle
gg|+\frac{1}{4}|ee\rangle\langle ee| \label{mixture}
\end{equation}
appears to be composed of the entangled state $\Psi_+$ with a
probability of 50\% and the two states $|gg\rangle$ and
$|ee\rangle$ with 25\% probability each. If the state $\Psi_+$ is
replaced by the density operator $\rho_p$ in the measurement
procedure described in subsection \ref{sec2:entangled}, the
resulting signal will be the same apart from a 50\% loss of
contrast. The states $|g\rangle|g\rangle$ and $|e\rangle|e\rangle$
do not contribute to the signal, since they become equally
distributed over all four basis states by the $\pi/2$ pulses
preceding the state detection. Their only effect is to reduce the
signal-to-noise ratio by adding quantum projection noise, since
only half of the experiments effectively contribute to the signal.

\begin{figure}
\resizebox{0.4\textwidth}{!}{%
  \includegraphics{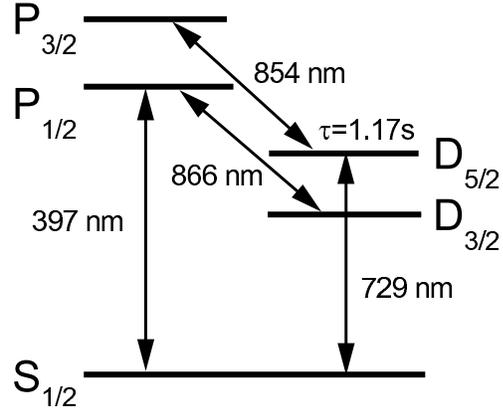} }
\caption{Relevant energy levels of $^{40}$Ca$^+$. For Doppler
cooling and quantum state detection, the ion is excited on the
transitions $S_{1/2}\leftrightarrow P_{1/2}\leftrightarrow
D_{3/2}$. Coherent operations are performed on the quadrupole
transition to the long-lived $D_{5/2}$ state. Population in the
$D_{5/2}$ state is pumped out via the $P_{3/2}$ level.}
\label{fig1:Ca40}       
\end{figure}

\section{Experimental setup}
\label{sec:1}
For our experiments, we confine strings of cold $^{40}$Ca$^+$ ions
in a linear ion trap. The radial confinement with trap frequencies
of about $\omega_\perp=(2\pi)\,4$~MHz is produced by a
radio-frequency quadrupole field oscillating at $23.5$~MHz. The
axial center-of-mass frequency $\omega_z$ is changed from 860~kHz
to 1720~kHz by varying the trap's tip voltages from 500 to 2000 V.
For these trapping parameters, the distance between two trapped
$^{40}$Ca$^+$ ions is between 3.9 $\mu$m and 6.2 $\mu$m.  Figure
\ref{fig1:Ca40} shows the lowest energy levels of $^{40}$Ca$^+$,
the most important feature being the metastable $D$ states with a
lifetime of about 1~s. The ions are Doppler cooled on the
$S_{1/2}\leftrightarrow P_{1/2}$ transition. Sideband cooling on
the $S_{1/2}\leftrightarrow D_{5/2}$ quadrupole transition can be
used to prepare a vibrational mode of the ion string in its ground
state. At the beginning of each experiment, the ions are
initialized in a pure state by optical pumping. Coherent quantum
state manipulation is achieved by laser pulses exciting a single
ion on the
$|S_{1/2},m\rangle\leftrightarrow|D_{5/2},m^\prime\rangle$
quadrupole transition. The Ti:sapphire laser used for exciting the
transition has a line width of below 50~Hz. Its strongly focused
beam is rapidly switched from one ion to the next by an
electro-optical deflector so that only one ion is interacting with
the beam at a time. We discriminate between the quantum states
$S_{1/2}$ and $D_{5/2}$ by scattering light on the
$S_{1/2}\leftrightarrow P_{1/2}$ dipole transition and detecting
the presence or absence of resonance fluorescence of the
individual ions with a CCD camera. For the measurement of operator
expectation values, the experiments are typically repeated one
hundred times. A more detailed account of the experimental setup
is given in Ref. \cite{Schmidtkaler03}.

\section{Applications}
\label{sec:2}
The measurement technique based on two unentangled atoms can be
used for frequency measurement as well as for the detection of
energy-level shifts. In the following subsections, it will be
applied to a measurement of magnetic field gradients, to a
determination of atomic electric quadrupole shifts and to a
measurement of the laser line width of a narrow-band laser.
\subsection{Detection of magnetic field gradients}
A magnetic field gradient pointing along the trap axis gives rise
to Zeeman shifts that are different for all the ions. This is
undesirable in experiments dedicated to quantum information
processing as it would necessitate the use of different laser
frequencies for performing the same type of operation on different
ions. In order to compensate residual magnetic field gradients
caused by the stray field of an ion getter pump and by
asymmetrically placed magnetic field coils, we measured the
magnetic field gradient by recording parity oscillations of a pair
of ions prepared in the state given in eq. (\ref{productstate})
with $|g\rangle\equiv|S_{1/2},m=1/2\rangle$ and
$|e\rangle\equiv|D_{5/2},m=5/2\rangle$.
As in this experiment the Zeeman shift is the only relevant
energy-level shift that is position dependent, the gradient was
minimized by applying a compensation gradient and minimizing the
parity oscillation frequency. For the chosen energy levels, the
dependence of the transition frequency on the magnetic field is
given by $d\nu/dB=2.8$~MHz/G. Ramsey probe times of several
hundred milliseconds make it possible to reduce the parity
oscillation frequency to below 1~Hz. This corresponds to a
difference of the magnetic fields at the location of the ions of
less than 0.4~$\mu$G and to a field gradient of below 0.08~G/m if
a typical ion distance of 5~$\mu$m is assumed. The precision of
the measurement could be further increased by operating the ion
trap at much lower axial trap frequencies in order to increase the
distance between the ions.
\subsection{Measurement of electric quadrupole shifts}
In optical frequency standards based on single hydrogen-like
trapped ions, the transition frequency from the $S$ ground state
to a metastable $D_j$ state is sensitive to electric field
gradients.
\begin{figure}
\centering
\resizebox{0.35\textwidth}{!}{%
  \includegraphics{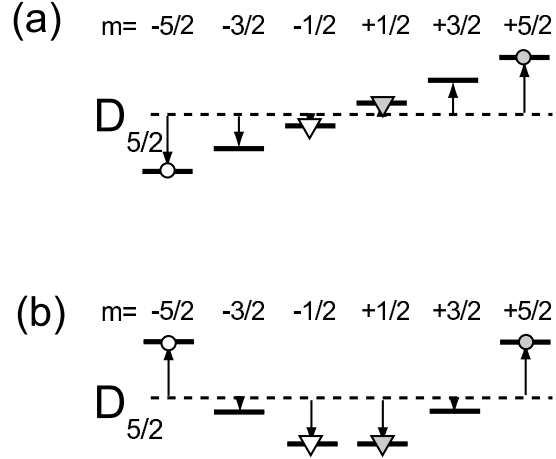}
}
\caption{Shift of the Zeeman substates of the metastable $D_{5/2}$
level by (a) a magnetic field and (b) an electric field gradient.
The magnetic quantum number is denoted by the symbol $m$. Energy
levels occurring in the entangled state
$\Psi_{ent}=\frac{1}{\sqrt{2}}
(|\circ\rangle|\bullet\rangle+|\triangledown\rangle|\blacktriangledown\rangle)$
 and the product state $\Psi_{pr}=\frac{1}{2}
(|\circ\rangle+|\triangledown\rangle)(|\bullet\rangle+|\blacktriangledown\rangle)$
 are denoted by the symbols
$\circ,\triangledown,\bullet,\blacktriangledown$, the open
(filled) symbols corresponding to levels occupied by atom 1 (2),
respectively. The combined Zeeman energy of the
$|\circ\rangle|\bullet\rangle$ states equals the energy of the
$|\triangledown\rangle|\blacktriangledown\rangle$- states, making
$\Psi_{ent}$ immune against magnetic field noise. The Bell state
and the product state used in the experiments (see
eqs.\,(\ref{Bellstate_exp}) and (\ref{productstate_exp})) have a
slightly smaller sensitivity to the quadrupole shift than the
states $\Psi_{ent}$ and $\Psi_{pr}$ but are easier to prepare.}
\label{fig2:levelshifts}       
\end{figure}
The $D_j$ state's atomic electric quadrupole moment interacting
with residual electric quadrupole fields\cite{Itano00} gives rise
to frequency shifts of a few hertz. The shift of the Zeeman
sublevel $|D_j,m_j\rangle$ in a quadrupole field $\Phi(x,y,z)=
A(x^2+y^2-2z^2)$ with rotational symmetry is given by
\begin{equation}
\label{shiftformula}
\hbar\Delta\nu=\frac{1}{4}\frac{dE_z}{dz}\Theta(D,j)\frac{j(j+1)-3{m_j}^2}{j(2j-1)}
(3\cos^2\beta-1),
\end{equation}
where $dE_z/dz=4A$ is the electric field gradient along the
potential's symmetry axis $z$, $\beta$ denotes the angle between
the potential's symmetry axis $\vec{e}_z$ and the magnetic field
vector $\vec{B}$, and $\Theta(D,j)$ expresses the strength of the
quadrupole moment in terms of a reduced matrix
element\cite{Itano00}. Quadrupole moments have been calculated
\cite{Itano06,Sur06} and experimentally determined
\cite{Oskay05,Barwood04,Schneider05} for $^{88}$Sr$^+$,
$^{199}$Hg$^+$ and $^{171}$Yb$^+$ by measuring the change in
transition frequency from the electronic ground state to the
metastable state that is affected by a static electric field
gradient. In principle, the shift could also be detected by
measuring transition frequencies between different Zeeman
sublevels. However, both measurement strategies become
impracticable in the presence of non-vanishing first-order Zeeman
shifts since residual fluctuating magnetic fields prevent the use
of sufficiently long laser-atom interaction times. In ref.
\cite{Roos06}, we demonstrated that this problem can be overcome
by preparing a pair of ions in an entangled state that is
decoherence free with respect to fluctuations of the magnetic
field but sensitive to the quadrupole shift. Fig.
\ref{fig2:levelshifts} shows the energy-level shifts of the
$D_{5/2}$ Zeeman sublevels under the influence of a magnetic field
and an electric quadrupole field. Labeling the Zeeman states
$|D_{5/2},m\rangle\equiv|m\rangle$ by their magnetic quantum
numbers, it can be seen that the Bell state
\begin{equation}
\Psi=\frac{1}{\sqrt{2}}(|\!-\!5/2\rangle|\!+\!3/2\rangle+|\!-\!1/2\rangle|\!-\!1/2\rangle)
\label{Bellstate_exp}
\end{equation}
is a superposition of two constituents having energies that shift
by the same amount in a magnetic field but shift in opposite
directions if an electric field gradient is applied. We used this
type of Bell state for a precise determination of the $D_{5/2}$
state's quadrupole moment in $^{40}$Ca$^+$ \cite{Roos06}.

\begin{figure}
\resizebox{0.45\textwidth}{!}{%
  \includegraphics{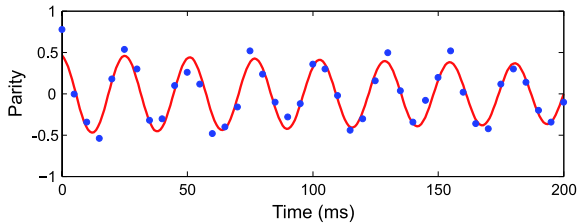}
}
\caption{Parity oscillations caused by the interaction of a static
electric field gradient with the quadrupole moment of the
$D_{5/2}$ state of $^{40}$Ca$^+$. The quadrupole shift is inferred
from the frequency of the parity oscillations. The first data
point significantly deviates from the fit since the quantum state
given by eq. (\ref{productstate_exp}) has not yet decayed to a
mixed quantum state. For the setting of the analysis pulses, the
pure state gives a significantly higher value of the parity.
Techniques similar to the ones later described in subsection {\it
\ref{subsec:laserlinewidth}} could in principle be used to avoid
this complication.}
\label{fig3:parityoscillation}       
\end{figure}
%

\begin{figure}
\resizebox{0.45\textwidth}{!}{%
    \includegraphics{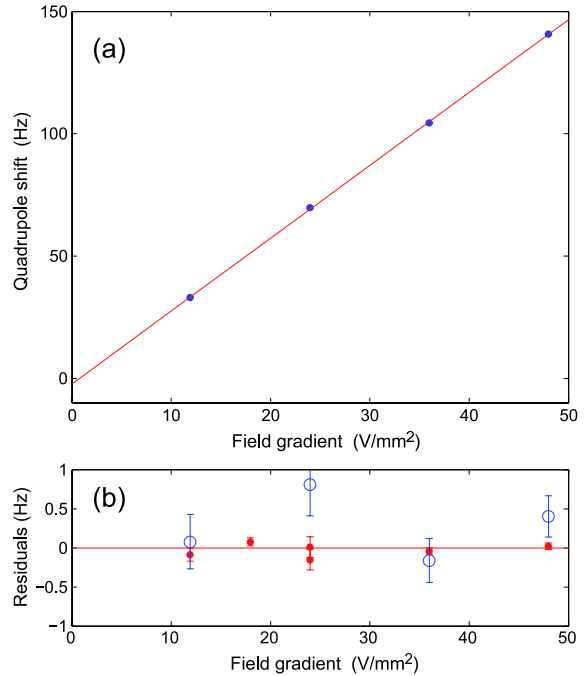}
}
\caption{Electric quadrupole shift measured with a pair of atoms
in a product state. (a) The shift varies linearly with the applied
electric field gradient. (b) Residuals of the electric quadrupole
shift measurements. The plot shows deviations of the data points
measured with unentangled ions (open circles) and entangled ions
(filled circles) with respect to the fit obtained from the
entangled state data. For the measurement with product states
(entangled states), $N\approx 20000$ ($N\approx 30000$) states
were prepared, waiting times of up to $100$~ms ($250$~ms) were
used, and the overall time of data taking was $6$ ($9$) h,
respectively.}
\label{fig4:quadrupoleshift}       
\end{figure}

To test whether the method outlined in subsection
\ref{sec2:unentangled} could be applied to a measurement of the
quadrupole moment, we prepared the state
\begin{equation}
\Psi_p=\frac{1}{2}(|\!-\!5/2\rangle+|\!-\!1/2\rangle)\otimes(|\!+\!3/2\rangle+|\!-\!1/2\rangle)
\label{productstate_exp}
\end{equation}
and let the fluctuating magnetic stray fields turn the pure state
into a mixed state within a few milliseconds. After a waiting time
ranging from 0.1 to 200 ms, $\pi/2$ pulses were applied and a
parity measurement performed. Fig. \ref{fig3:parityoscillation}
shows the resulting parity oscillations whose contrast decays over
a time interval orders of magnitude longer than any single-atom
coherence time in $^{40}$Ca$^+$. A sinusoidal fit to the data
reveals an initial contrast of 48(6)\% and an oscillation
frequency $\nu=38.6(3)$~Hz. For the fit, the first data point at
t=0.1$\mu$s is not taken into account. At this time, the quantum
state cannot yet be described by a mixture similar to the one of
eq.~(\ref{mixture}) as some of the coherences persist for a few
milliseconds and thus affect the parity signal. The parity signal
decays exponentially with a time constant $\tau_d=730(530)$~ms
that is consistent with the assumption of spontaneous decay being
the only source of decoherence (in this case, one would have
$\tau_d=\tau_{D_{5/2}}/2\approx 580$~ms, where $\tau_{D_{5/2}}$ is
the lifetime of the metastable state). The quadrupole moment is
determined by measuring the quadrupole shift as a function of the
electric field gradient $E^\prime$. The latter is conveniently
varied by changing the voltage applied to the axial trap
electrodes. For a calibration of the gradient, the axial
oscillation frequency of the ions is measured. Further details
regarding the measurement procedure are provided in ref.
\cite{Roos06}. Fig. \ref{fig4:quadrupoleshift}(a) shows the
quadrupole shift $\Delta\nu_{QS}$ as a function of the field
gradient (the small offset at $E^\prime=0$ is caused by the
second-order Zeeman effect). By fitting a straight line to the
data, the quadrupole moment can be calculated provided that the
angle between the orientation of the electric field gradient and
the quantization axis is known. Setting $\Delta\nu_{QS}=\alpha
E^\prime$, the fit yields the proportionality constant
$\alpha=2.977(11)$ Hz/(V/mm$^2$). 
For the same configuration, the quadrupole shift was also measured
with an entangled state, giving $\alpha=2.975(2)$ Hz/(V/mm$^2$)
and thus confirming that the measurements with product and with
entangled states give the same result.
Fig.~\ref{fig4:quadrupoleshift}(b) shows the deviations of the
data points measured with entangled and with classically
correlated ions with respect to the fit obtained from the
entangled state data. As expected, the error bars of the
measurement with classically correlated ions are bigger since the
Ramsey contrast is reduced and the quantum projection noise is
higher. A rough estimation of the influence of quantum projection
noise on the uncertainty $\sigma$ of the quadrupole shift
$\Delta\nu_{QS}$ is obtained by setting
$\sigma_{\Delta\nu_{QS}}\backsimeq 1/((\tau/2)(C\sqrt{2N}))$,
where $\tau$ is the maximum delay between state preparation and
parity measurement, $C$ the resulting Ramsey contrast, and $N$
denotes the number of experiments (for two ions, $2N$ state
measurements contribute to the signal). For the experiments with
entangled states shown in Fig.~\ref{fig4:quadrupoleshift}b, where
$N\approx 30000$, $C\approx 0.5$, and $\tau=250$~ms, one expects
$\sigma_{\Delta\nu_{QS}}\approx 65$~mHz. For the product state
experiments ($N\approx 20000$, $C\approx 0.3$, $\tau=100$~ms), one
expects $\sigma_{\Delta\nu_{QS}}\approx 330$~mHz. These estimates
are consistent with the experimental observations. Even though
experiments with product states suffer from a reduced Ramsey
contrast, they are attractive because these experiments are
considerably easier to realize. Contrary to the ones based on
entangled states, they do not require cooling the ions to the
ground state of the trapping potential by sideband cooling on the
quadrupole transition using a laser with a narrow line width.
Moreover, it turns out that the precision of the quadrupole moment
measurement is not limited by the statistical errors but rather by
a number of systematic errors, the most important one being the
alignment of the magnetic field vector defining the quantization
axis with the symmetry axis of the trap \cite{Roos06} described by
the angle $\beta$ in eq. (\ref{shiftformula}). As can be seen from
(\ref{shiftformula}), any misalignment will reduce the quadrupole
shift and thus result in an underestimation of the quadrupole
moment. In ref.~\cite{Roos06}, the precision of the measurement
was limited by the estimated uncertainty of $\Delta\beta=3^\circ$.

\subsection{Laser line width
measurements\label{subsec:laserlinewidth}}
The linewidth of a narrow-band laser is often investigated by
setting up two identical laser systems and observing their beat
signal. If no second laser system is available, Ramsey experiments
are a suitable tool for determining the laser line width
\cite{Sengstock94}. This, however, requires the atomic transition
line width not to be broadened by phase noise to an extent
comparable to the laser line width to be measured. In
$^{40}$Ca$^+$, this requirement is difficult to fulfil for lasers
with a line width of 100 Hz or below as all transitions depend on
the magnetic field in first order. Here again, a solution consists
in performing Ramsey measurements with a pair of ions in a
correlated product state. For a measurement of the line width of
the Ti:Sapphire laser exciting the $^{40}$Ca$^+$ ions on the
$S_{1/2}\leftrightarrow D_{5/2}$ transition, we prepared the state
$\Psi_p=\psi_1\otimes\psi_2$ where
\begin{eqnarray*}
\psi_1&=&\frac{1}{\sqrt{2}}(|S_{1/2},m=\!-1/2\rangle+|D_{5/2},m=\!-1/2\rangle)\\
\psi_2&=&\frac{1}{\sqrt{2}}(|S_{1/2},m=\!+1/2\rangle+|D_{5/2},m=\!+1/2\rangle)
\end{eqnarray*}
denote the states of ion 1 and ion 2. The fringes of the
single-ion Ramsey signals
$\langle\sigma_z^{(n)}\rangle=\cos(\phi_L+(-1)^n\phi_B+\phi_n)$,
$(n=1,2)$, show the same dependence on changes of the laser
frequency but shift in opposite directions under changes of the
magnetic field. Here, $\phi_L$ is the phase shift arising from
deviations of the laser frequency from the atomic transition
frequency, $\phi_B$ denotes shifts due to deviations of the
magnetic field from its mean value, and $\phi_n$ is the phase
difference between the first and the second Ramsey pulses. Then,
the two-ion parity signal is given by
\begin{eqnarray}
\label{paritylaser}
\langle\sigma_z^{(1)}\sigma_z^{(2)}\rangle&=&\langle\cos(\phi_L\!-\!\phi_B\!+\!\phi_1)\cos(\phi_L\!+\!\phi_B\!+\!\phi_2)\rangle\\
&=&\frac{1}{2}(\langle\cos(2\phi_L\!+\!\phi_2\!+\!\phi_1)\rangle\!+
\!
\langle\cos(2\phi_B\!+\!\phi_2\!-\!\phi_1)\rangle)\nonumber
\end{eqnarray}
In the parity signal, the phase noise contributions from the laser
and the magnetic field separate. Moreover, it is even possible to
make either the first or the second term vanish by the following
procedure: for an estimation of the expectation value of the
observable $\sigma_z^{(1)}\sigma_z^{(2)}$, a Ramsey experiment is
carried out $N$ times. If the phases $\phi_1,\phi_2$ are chosen to
be random variables by setting $\phi_1=\phi_{0}+\phi_X$,
$\phi_2=-\phi_X$, where $\phi_X$ is a random variable with uniform
distribution over the interval $[0,2\pi)$, the
magnetic-field-dependent term averages to zero and
(\ref{paritylaser}) becomes
\begin{equation}
\langle\sigma_z^{(1)}\sigma_z^{(2)}\rangle=
\frac{1}{2}\langle\cos(2\phi_L\!+\!\phi_0)\rangle
\end{equation}
as only the sum $\phi_1+\phi_2=\phi_0$ is not a random number. By
setting $\phi_1=\phi_{0}+\phi_X$, $\phi_2=\phi_X$, the experiment
could also be made sensitive to decoherence caused by magnetic
field noise and insensitive to laser frequency noise. In both
cases, all single-ion coherences are completely averaged out. By
scanning the phase $\phi_0$, Ramsey fringes are recorded for the
parity $\sigma_z^{(1)}\sigma_z^{(2)}$ and the resulting Ramsey
contrast can be plotted as a function of the Ramsey time. Note
that the Ramsey contrast will decay twice as fast as in a standard
Ramsey experiment with a single atom because of the factor
$2\phi_L$ in eq.(\ref{paritylaser}).

\begin{figure}
\resizebox{0.45\textwidth}{!}{%
    \includegraphics{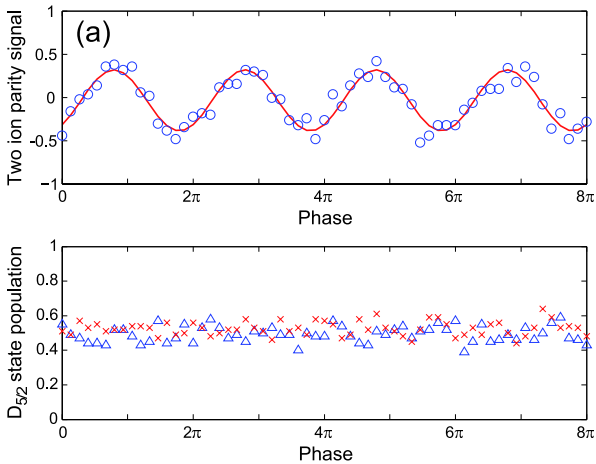}
}
\caption{Two-ion Ramsey experiment for a measurement of the laser
line width. (a) Two-ion parity signal measured as a function of
$\phi_0$ for a Ramsey time $\tau=1.51$~ms. (b) Single-ion
coherences. Because of the randomization of the second pulses'
phase with respect to the phase of the first pulse, the signal is
completely flat.}
\label{fig6:TwoIonPhaseScan}       
\end{figure}
\begin{figure}
\resizebox{0.45\textwidth}{!}{%
  \includegraphics{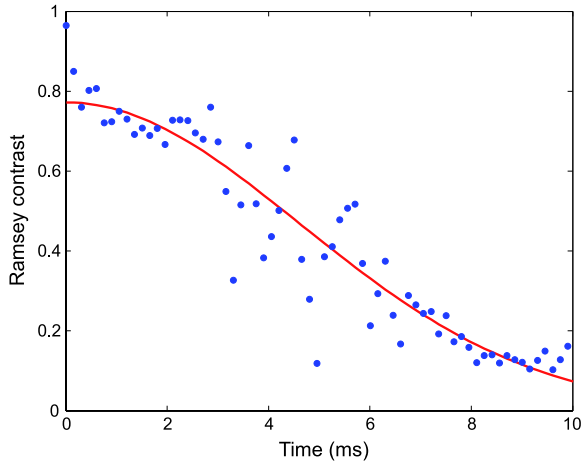}
} \caption{Measurement of the laser linewidth by a two-ion Ramsey
experiment. The Ramsey contrast as a function of time is fitted by
a Gaussian with half-width $\tau_{1/2}=4.6(2)$~ms.}
\label{fig7:RamseyContrastWithProductStates}       
\end{figure}
Ramsey fringes recorded by this technique are shown in Fig.
\ref{fig6:TwoIonPhaseScan}. The figure's upper panel displays the
two-ion parity signal as a function of $\phi_0$ for a Ramsey time
$\tau=1.51$~ms. The lower panel shows the single-ion signals
$\langle \sigma_z^{(1)}\rangle$, $\langle \sigma_z^{(2)}\rangle$,
having no contrast at all. In Fig.
\ref{fig7:RamseyContrastWithProductStates}, the two-ion Ramsey
contrast is plotted as a function of the Ramsey time $\tau$. As
low-frequency laser noise is the dominant source of noise
broadening the laser line width in our experiments, the Ramsey
contrast data are fitted by a Gaussian function with a half-width
$\tau_{1/2}=4.6(2)$~ms. From this value, we infer a laser line
width of 48(2)~Hz on a time scale of one minute.

\section{Conclusions}

In ion-trap experiments, one is often confronted with collective
decoherence mechanisms caused by external fluctuating fields with
spatial correlation lengths much longer than the typical distance
between ions. In these conditions, experiments with correlated
ions prepared in decoherence-free subspaces allow for
substantially longer coherence times than experiments
investigating single-ion coherences only. While the preparation of
entangled states is optimum for achieving high signal-to-noise
ratios, substantial improvements can already be achieved by
experimenting with ions that are classically correlated but not
entangled.

\begin{acknowledgement}
We acknowledge support by the Austrian Science Fund (FWF), the
European Commission (SCALA, CONQUEST networks), and by the
Institut f\"ur Quanteninformation GmbH. K.~K acknowledges funding
by the Lise Meitner program of the FWF. C.~F.~R would like to
thank Bodensteinalm for hospitality during the last stage of the
preparation of the manuscript.
\end{acknowledgement}


\begin{thebibliography}{}
%
%
\bibitem{Margolis04} H.~S.~Margolis, G.~P.~Barwood, G.~Huang, H.~A.~Klein,S.~N.~Lea, K.~Szymaniec, P.~Gill, Science \textbf{306}, 1355 (2004)

\bibitem{Schneider05} T.~Schneider, E.~Peik, C.~Tamm, Phys.~Rev.~Lett. \textbf{94}, 230801 (2005)

\bibitem{Oskay06} W.~H.~Oskay, S.~A.~Diddams, E.~A.~Donley, T.~M.~Fortier, T.~P.~Heavner, L.~Hollberg, W.~M.~Itano, S.~R.~Jefferts, M.~J.~Delaney, K.~Kim, F.~Levi, T.~E.~Parker, J.~C.~Bergquist, Phys.~Rev.~Lett. \textbf{97}, 020801 (2006)

\bibitem{Rosenband07} T.~Rosenband, P.~O.~Schmidt, D.~B.~Hume, W.~M.~Itano, T.~M.~Fortier, J.~E.~Stalnaker, K.~Kim, S.~A.~Diddams, J.~C.~J.~Koelemeij, J.~C.~Bergquist, D.~J.~Wineland, Phys.~Rev.~Lett. \textbf{98}, 220801 (2007)

\bibitem{Leibfried05} D.~Leibfried, E.~Knill, S.~Seidelin, J.~Britton, R.~B.~Blakestad, J.~Chiaverini, D.~B.~Hume, W.~M.~Itano, J.~D.~Jost, C.~Langer, R.~Ozeri, R.~Reichle, D.~J.~Wineland, Nature \textbf{438}, 639 (2005)

\bibitem{Haeffner05} H.~H\"affner, W.~H\"ansel, C.~F.~Roos, J.~Benhelm, D.~Chek-al-kar, M.~Chwalla, T.~K\"orber, U.~D.~Rapol, M.~Riebe, P.~O.~Schmidt,
C.~Becher, O.~G\"uhne, W.~D\"ur, R.~Blatt, Nature \textbf{438},
643 (2005)

\bibitem{Riebe04} M.~Riebe, H.~H\"affner, C.~F.~Roos, W.~H\"ansel, J.~Benhelm, G.~P.~T.~Lancaster, T.~W.~K\"orber, C.~Becher, F.~Schmidt-Kaler, D.~F.~V.~James, R.~Blatt, Nature \textbf{429}, 734 (2004)

\bibitem{Barrett04} M.~D.~Barrett, J.~Chiaverini, T.~Sch\"atz, J.~Britton, W.~M.~Itano, J.~D.~Jost, E.~Knill, C.~Langer, D.~Leibfried, R.~Ozeri, D.~J.~Wineland, Nature \textbf{429}, 737 (2004)

\bibitem{Bollinger96} J.~J.~Bollinger, W.~M.~Itano, D.~J.~Wineland, D.~J.~Heinzen,
Phys.~Rev.~A \textbf{54}, R4649 (1996)

\bibitem{Leibfried04} D.~Leibfried, M.~D.~Barrett, T.~Sch\"atz, J.~Britton, J.~Chiaverini, W.~M.~Itano, J.~D.~Jost, C.~Langer, D.~J.~Wineland, Science \textbf{304}, 1476 (2004)

\bibitem{Schmidt05} P.~O.~Schmidt, T.~Rosenband, C.~Langer,
W.~M.~Itano, J.~C.~Bergquist, D.~J.~Wineland, Science
\textbf{309}, 749 (2005)

\bibitem{Lidar98} D.~A.~Lidar, I.~L.~Chuang, K.~B.~Whaley, Phys.~Rev.~Lett. \textbf{81}, 2594 (1998)

\bibitem{Kielpinski01} D.~Kielpinski, V.~Meyer, M.~A.~Rowe, C.~A.~Sackett, W.~M.~Itano, C.~Monroe, D.~J.~Wineland, Science \textbf{291}, 1013 (2001)

\bibitem{Roos04} C.~F.~Roos, G.~P.~T.~Lancaster, M.~Riebe, H.~H\"affner, W.~H\"ansel, S.~Gulde, C.~Becher, J.~Eschner, F.~Schmidt-Kaler, R.~Blatt, Phys.~Rev.~Lett.
\textbf{92}, 220402 (2004)

\bibitem{Roos05} C.~F.~Roos, arXiv:quant-ph/0508148 (2005)

\bibitem{Roos06}
C.~F.~Roos, M.~Chwalla, K.~Kim, M.~Riebe, R.~Blatt, Nature
\textbf{443}, 316 (2006)

\bibitem{Schmidtkaler03} F.~Schmidt-Kaler, H.~H\"affner, S.~Gulde, M.~Riebe, G.~P.~T.~Lancaster, T.~Deuschle, C.~Becher, W.~H\"ansel, J.~Eschner, C.~F.~Roos, R.~Blatt, Appl. Phys. B \textbf{77}, 789 (2003)

\bibitem{Itano00} W.~M.~Itano, J.~Res.~Natl.~Inst.~Stand.~Technol.
\textbf{105}, 829 (2000)

\bibitem{Itano06} W.~M.~Itano, Phys.~Rev.~A \textbf{73}, 022510 (2006)

\bibitem{Sur06}  C.~Sur, K.~V.~P.~Latha, B.~K.~Sahoo, R.~K.~Chaudhuri, B.~P.~Das, Phys.~Rev.~Lett. \textbf{96}, 193001 (2006)

\bibitem{Oskay05} W.~H.~Oskay, W.~M.~Itano, J.~C.~Bergquist,
Phys.~Rev.~Lett. \textbf{94}, 163001 (2005)

\bibitem{Barwood04} G.~P.~Barwood, H.~S.~Margolis, G.~Huang, P.~Gill, H.~A.~Klein,
Phys.~Rev.~Lett. \textbf{ 93}, 133001 (2004)

\bibitem{Sengstock94} K.~Sengstock, U.~Sterr, J.~H.~M{\"u}ller, V.~Rieger, D.~Bettermann,
W.~Ertmer, Appl.~Phys.~B \textbf{59}, 99 (1994)


\end{thebibliography}
%

\end{document}